\documentclass[preprint,12pt]{elsarticle}

\usepackage{graphicx}

\usepackage{amssymb}
\usepackage{amsthm}
\usepackage{amsmath}

\usepackage{here}

\makeatletter
\def\simleq{\mathrel{\mathpalette\gl@align<}}
\def\simgeq{\mathrel{\mathpalette\gl@align>}}
\def\gl@align#1#2{\lower.6ex\vbox{\baselineskip\z@skip\lineskip\z@
     \ialign{$\m@th#1\hfill##\hfil$\crcr#2\crcr\sim\crcr}}}
\makeatother

\newcommand{\nn}{\nonumber\\}

\newcommand{\scsize}[1]{{\scriptsize #1}}

\begin{document}

\begin{frontmatter}



\title{Unified contraction algorithm for multi-baryon correlators on the lattice}


\author{Takumi~Doi\corref{cor1}}
\cortext[cor1]{Corresponding author}
\ead{doi@ribf.riken.jp}

\author{Michael G. Endres}

\address
{Theoretical Research Division, Nishina Center, RIKEN, Wako 351-0198, Japan}

\begin{abstract}
We propose a novel algorithm for calculating multi-baryon correlation functions on the lattice.
By considering the permutation of quarks (Wick contractions) 
and color/spinor contractions simultaneously, 
we construct a unified index list for the contraction where the redundancies 
in the original contraction are eliminated.
We find that a significant reduction in the computational
cost of correlators is achieved, e.g., 
by a factor of 192 for $^3$H and $^3$He nuclei, and 
a factor of 20736 for the $^4$He nucleus,
without assuming isospin symmetry.
A further reduction is possible by exploiting isospin symmetry, 
and/or interchange symmetries associated with sink baryons, if such symmetries exist.
Extensions for systems with hyperons are presented as well.
\end{abstract}


\end{frontmatter}



\section{Introduction}
\label{sec:intro}

Correlation functions of multi-baryon systems are the central quantities
to be calculated
when determining the properties and interactions of atomic nuclei
directly from lattice QCD (+QED) simulations.
The computational cost of constructing such correlators is, however, known to be exceptionally enormous
for large mass number $A$,
and
one of the greatest challenges is to find an efficient algorithm for reducing it.
The reason for such a high cost is that (i) the number of quark permutations (Wick contractions) grows  
factorially with $A$
and
(ii) the contraction of color/spinor degrees of freedom (DoF) becomes 
exponentially large for large $A$.
While there has been significant progress toward reducing this computational 
cost~\cite{Ishii:2006ec, Ishii:2009zr, Aoki:2009ji, Yamazaki:2009ua, Doi:2011gq},
it continues to remain the most time-consuming part of the calculation,
particularly for $A > 2$.%
\footnote{
In this paper, we consider the computational cost of constructing 
multi-baryon correlators for a fixed ensemble size, neglecting 
the potential computational difficulty of achieving an acceptable signal/noise 
ratio for such correlators at late times.
This unrelated issue, known as the signal/noise problem, 
grows exponentially with both mass number and time separation of correlators~\cite{Lepage:1989hd}.
For various attempts to ameliorate this problem, see, e.g., 
Refs.~\cite{Luscher:1990ck,Fleming:2004hs,Beane:2009kya,Beane:2009gs,Yamazaki:2011nd,Ishii:2012aa}.
}

Lattice QCD simulations for multi-baryon systems date back to Refs.~\cite{Fukugita:1994na, Fukugita:1994ve},
where energies of two-nucleon (2N) systems in a box were extracted 
from temporal correlators in Euclidean space-time, 
and then related to 2N scattering lengths through 
the use of L\"{u}scher's formula~\cite{Luscher:1985dn, Luscher:1986pf, Luscher:1990ux}.
Similar methods have been employed 
in recent studies as well~\cite{Yamazaki:2011nd, Beane:2011iw}.
In Refs.~\cite{Ishii:2006ec, Ishii:2009zr, Aoki:2009ji}, 
a new approach 
had been proposed,  where 
nuclear forces were directly extracted 
from Nambu-Bethe-Salpeter (NBS) wave functions, or spacial correlators of 2N systems.
This method was successfully extended to general hadron-hadron interactions 
such as hyperon-nucleon (YN) and hyperon-hyperon (YY) potentials%
~\cite{Nemura:2008sp, Inoue:2010hs, Inoue:2010es, Murano:2011nz, Aoki:2011gt, Inoue:2011ai, Sasaki:2011AAA, Ikeda:2011qm}.
A further extension was proposed in Ref.~\cite{Ishii:2012aa}, 
where both spacial and temporal dependencies of correlators are utilized
to extract non-local hadron-hadron potentials without requiring ground state saturation.%

Only quite recently, however, 
have lattice QCD studies for three- and higher-baryon systems been initiated:
the NPLQCD Collaboration demonstrated a feasibility study for the energy 
of a system with $\Xi^0\Xi^0n$ quantum numbers~\cite{Beane:2009gs};
the PACS-CS Collaboration studied the energies of $^3$He and $^4$He at several lattice volumes and 
concluded that both $^3$He and $^4$He are bound states~\cite{Yamazaki:2009ua}.
In Ref.~\cite{Doi:2011gq}, the HAL QCD Collaboration investigated 
three-nucleon forces (3NF) using the NBS wave function of three nucleons,
and repulsive 3NF at short distance were found in the triton ($^3$H) channel.
One of the major obstacles in each of these studies was 
the computational cost of quark contractions, as discussed above.

The purpose of this paper is to present a novel algorithm for the computation of
multi-baryon correlators.
In particular, by considering 
the quark permutation and
the color/spinor contractions 
simultaneously,
we show that there exist large redundancies in the contributions to the correlator.
By constructing a unified index list of non-vanishing contributions to the contraction
with those redundancies eliminated,
we can achieve a significant speedup for the computation of correlators.
As will be described later,
it is not necessary to assume a symmetry between different flavors (e.g., isospin symmetry)
in this algorithm, although by doing so, an additional reduction in cost can be achieved.

This paper is organized as follows.
In Section~\ref{sec:corr},
we describe the multi-baryon correlation functions under consideration
and review the issues associated with computing contractions.
In Section~\ref{sec:unified}, we propose a new algorithm
which utilizes a unified index list for evaluating contractions.
In Section~\ref{sec:results}, the efficiency of the new algorithm is discussed, while 
Section~\ref{sec:summary} is devoted to summary and concluding remarks.
Further details of our results are tabulated in~\ref{sec:tabs}.

\section{Multi-baryon correlation functions}
\label{sec:corr}

We consider a $2A$-point multi-baryon correlation function 
with a mass number $A$, defined by
\begin{eqnarray}
\label{eq:corr}
\lefteqn{
\Pi_{
\alpha_1, \cdots,\alpha_A;\ 
\alpha'_1,\cdots,\alpha'_A} 
(X_1,\cdots,X_A;\ X'_1,\cdots,X'_A)
} \nn
&\equiv&
\langle
B_{\alpha_1}(X_1) \cdots B_{\alpha_A}(X_A) 
\bar B'_{\alpha'_A}(X'_A) \cdots \bar B'_{\alpha'_1}(X'_1) 
\rangle ,
\end{eqnarray}
where 
$B_{\alpha_i} (\bar B'_{\alpha'_i})$ denotes an appropriate 
baryon interpolating field in the sink (source) with
a spinor index $\alpha_i$ ($\alpha'_i$), and
coordinate index $X_i \equiv (t_i, \vec{X}_i)$ ($X'_i$). 
%
We consider a general baryon operator given by
\begin{eqnarray}
B_{\alpha} (X) &=& 
\epsilon_{c_1 c_2 c_3} (C \Gamma_1)_{\alpha_1,\alpha_2} (\Gamma_2)_{\alpha,\alpha_3} q(\xi_1) q(\xi_2) q(\xi_3) , \\
\bar{B'}_{\alpha'} (X') &=& 
\epsilon_{c'_1 c'_2 c'_3} (C \Gamma'_1)_{\alpha'_1,\alpha'_2} (\Gamma'_2)_{\alpha',\alpha'_3} \bar{q}(\xi'_3) \bar{q}(\xi'_2) \bar{q}(\xi'_1) ,
\end{eqnarray}
where $c_i$ ($c'_i$) denotes color indices, and
$\xi_i$ ($\xi'_i$) is a symbolic label for the collection of indices $\{x_i, c_i, \alpha_i\}$ with $x_i$ being a quark coordinate index. 
Summation over repeated indices is implied.
$C=\gamma_4\gamma_2$ is the charge conjugation matrix and
$\Gamma_i$ ($\Gamma'_i$) are appropriate $\gamma$-matrices.
For instance, the choice of $(\Gamma_1, \Gamma_2) = (\Gamma'_1, \Gamma'_2) = (\gamma_5, 1)$ is often employed 
for an octet baryon field.
%
In the case of point sources and point sinks for quark fields, $X = x_1 = x_2 = x_3$ and $X' = x'_1 = x'_2 = x'_3$.
Generalization to smeared quark fields in the sink and/or source is straight-forward; in such cases the coordinate indices $x_i$ ($x'_i$) are replaced by associated smearing parameters.

\subsection{Computation using a straightforward algorithm}
\label{subsec:naive}

As described in Sec.~\ref{sec:intro}, 
the computational cost of a multi-baryon correlator 
diverges quickly
for large $A$.
One can estimate the number of contractions 
in Eq.~(\ref{eq:corr}) for a given $\{\alpha_i,\alpha'_i,t_i,t'_i\}$ as follows.
First, if we consider the 3-flavor space,
the number of quark permutations (Wick contractions) amount to
$N_{\rm perm} = N_u ! \cdot N_d ! \cdot N_s! $
where $N_u, N_d$ and $N_s$ are the number of up, down and strange quarks in the system, respectively.
For instance, $N_{\rm perm} = 36$ for $^2$H, 2880 for $^3$H/$^3$He and 518400 for $^4$He.
Second, one must take into account the contractions for the color/spinor DoF.
To carry out the counting, we exploit the sparse nature of $\gamma$-matrices and $\epsilon$-tensor:
for each baryon in the sink or source, we attribute a factor of six to each color loop (i.e., sum over each color index),
and a factor of four to each spinor loop (i.e., sum over each spinor index).
The total cost of the color/spinor contractions therefore scale as $N_{\rm loop} = 6^{2A}\cdot 4^{2A}$.
Particularly, we find $N_{\rm loop}$ to be ${\cal O}(10^5)$ for $^2$H, 
${\cal O}(10^8)$ for $^3$H/$^3$He and ${\cal O}(10^{11})$ for $^4$He.
Third, we must repeat the above computation for 
all possible spacial variables at the sink, $\{\vec{X}_1,\cdots,\vec{X}_A\}$.
For instance, when extracting the energy of the system,
a zero-momentum projection for each baryon is commonly performed%
~\cite{Yamazaki:2009ua, Beane:2009gs, Yamazaki:2011nd, Fukugita:1994na, Fukugita:1994ve, Beane:2011iw}.
When determining hadron-hadron potentials, 
NBS wave functions are extracted by 
imposing zero-momentum for the center of gravity,
and the dependencies on relative coordinates between baryons are subject of interest%
~\cite{Ishii:2006ec, Ishii:2009zr, Aoki:2009ji, Doi:2011gq, Ishii:2012aa,
Nemura:2008sp, Inoue:2010hs, Inoue:2010es, Murano:2011nz, Aoki:2011gt, Inoue:2011ai, Sasaki:2011AAA, Ikeda:2011qm}.
In both cases, the computational cost will be multiplied by a factor of $N_{\rm vol} = L^{3A}$,
where $L$ is the spacial extent of the lattice.%
\footnote{
One typically does not count the computational cost associated with summing over
spacial baryon coordinates, $\{\vec X'_1,\cdots,\vec X'_A\}$, at the source.
Rather, a sum over quark coordinate indices is implicitly performed at the source by solving quark propagators with smearing sources.
If one uses, e.g., all-to-all propagators, an additional factor of $N_{\rm vol}$ would arise, however.
}

\subsection{Block algorithm}
\label{subsec:block}

Recently, algorithmic
progress has been achieved for the computation 
of multi-baryon correlators~\cite{Ishii:2006ec, Ishii:2009zr, Aoki:2009ji, Yamazaki:2009ua, Doi:2011gq}, by considering
a block of three-quark propagators combined into a baryon sink.
For simplicity, let us consider a $2A$--point nucleon correlation function with $A=2$, given by
\begin{eqnarray}
\label{eq:pn}
\Pi_{
\alpha, \beta;\
\alpha',\beta'}
(X_1,X_2;\ X'_1,X'_2)
&=&
\langle
p_{\alpha}(X_1) n_{\beta}(X_2)
\bar n'_{\beta'}(X'_2) \bar p'_{\alpha'}(X'_1) 
\rangle ,
\end{eqnarray}
where the proton and neutron fields are defined by
\begin{eqnarray}
p_{\alpha} (X) &=& 
+ \epsilon_{c_1 c_2 c_3} (C \Gamma^p_1)_{\alpha_1,\alpha_2} (\Gamma^p_2)_{\alpha,\alpha_3} u(\xi_1) d(\xi_2) u(\xi_3) , \\
n_{\alpha} (X) &=& 
- \epsilon_{c_1 c_2 c_3} (C \Gamma^n_1)_{\alpha_1,\alpha_2} (\Gamma^n_2)_{\alpha,\alpha_3} d(\xi_1) u(\xi_2) d(\xi_3) , \\
\bar{p'}_{\alpha'} (X') &=& 
+ \epsilon_{c'_1 c'_2 c'_3} (C \Gamma'^{\,p}_1)_{\alpha'_1,\alpha'_2} (\Gamma'^{\,p}_2)_{\alpha',\alpha'_3} \bar{u}(\xi'_3) \bar{d}(\xi'_2) \bar{u}(\xi'_1) , \\
\bar{n'}_{\alpha'} (X') &=& 
- \epsilon_{c'_1 c'_2 c'_3} (C \Gamma'^{\,n}_1)_{\alpha'_1,\alpha'_2} (\Gamma'^{\,n}_2)_{\alpha',\alpha'_3} \bar{d}(\xi'_3) \bar{u}(\xi'_2) \bar{d}(\xi'_1) .
\end{eqnarray}
We construct blocks of three-quark propagators defined by
\begin{eqnarray}
\label{eq:fp}
\lefteqn{
f_\alpha^p(X;\ \xi'_1, \xi'_2, \xi'_3)
\equiv
\langle p_\alpha (X) \cdot \bar{u}(\xi'_3) \bar{d}(\xi'_2) \bar{u}(\xi'_1) \rangle
} \nn
&=& \epsilon_{c_1 c_2 c_3} (C \Gamma^p_1)_{\alpha_1,\alpha_2} (\Gamma^p_2)_{\alpha,\alpha_3} \nn
&& 
\times\left[\ S_u(\xi_1,\xi'_1) S_u(\xi_3,\xi'_3) - S_u(\xi_1,\xi'_3) S_u(\xi_3,\xi'_1)\ \right]
S_d(\xi_2,\xi'_2) ,
\end{eqnarray}
and
\begin{eqnarray}
\label{eq:fn}
\lefteqn{
f^n_\beta (X;\ \xi'_1, \xi'_2, \xi'_3)
\equiv
- \langle n_\beta(X) \cdot \bar{d}(\xi'_3) \bar{u}(\xi'_2) \bar{d}(\xi'_1) \rangle
} \nn
&=& \epsilon_{c_1 c_2 c_3} (C \Gamma^n_1)_{\alpha_1,\alpha_2} (\Gamma^n_2)_{\beta,\alpha_3} \nn
&&
\times\left[\ S_d(\xi_1,\xi'_1) S_d(\xi_3,\xi'_3) - S_d(\xi_1,\xi'_3) S_d(\xi_3,\xi'_1)\ \right]
S_u(\xi_2,\xi'_2) ,
\end{eqnarray}
for all possible indices $\{X, \xi'_i\}$,
where $S_q(\xi_i,\xi'_j) \equiv \langle q(\xi_i) \bar{q}(\xi'_j) \rangle$ denotes a quark propagator associated
with the flavor $q=\{u,d,s\}$.
Using Eqs.~(\ref{eq:fp}) and (\ref{eq:fn}),
the correlation function can be written as
\begin{eqnarray}
\label{eq:block}
\lefteqn{
\Pi_{
\alpha, \beta;\
\alpha',\beta'}
(X_1,X_2;\ X'_1,X'_2)
} \nn
&=& 
\sum_{\sigma} 
f_\alpha^p(X_1;\ \xi'_{\sigma(1)}, \xi'_{\sigma(2)}, \xi'_{\sigma(3)}) 
\cdot
f_\beta^n(X_2;\ \xi'_{\sigma(4)}, \xi'_{\sigma(5)}, \xi'_{\sigma(6)}) \nn
&& 
\times
\epsilon_{c'_1 c'_2 c'_3} (C \Gamma'^{\,p}_1)_{\alpha'_1,\alpha'_2} (\Gamma'^{\,p}_2)_{\alpha',\alpha'_3} 
\cdot
\epsilon_{c'_4 c'_5 c'_6} (C \Gamma'^{\,n}_1)_{\alpha'_4,\alpha'_5} (\Gamma'^{\,n}_2)_{\beta',\alpha'_6}
\cdot
{\rm sign}(\sigma) , \ \ \ 
\end{eqnarray}
where $\sum_\sigma \equiv \sum_{\sigma_u} \sum_{\sigma_d}$ with 
$\sigma_u$ ($\sigma_d$) representing the permutation among up (down) quarks,
and ${\rm sign}(\sigma) = {\rm sign}(\sigma_u) {\rm sign}(\sigma_d)$ representing a sign factor which arises from the anti-commuting property of fermions.

There are several significant advantages to using Eq.~(\ref{eq:block}) over the straightforward approach.
First, in terms of the permutation, we note that 
Eq.~(\ref{eq:fp}) is antisymmetric under the exchange of two up quarks in the proton.
A similar property holds for Eq.~(\ref{eq:fn}) under the exchange of two down quarks in the neutron.
Generally speaking, by exploiting these features, one can restrict the full permutation appearing in
Eq.~(\ref{eq:block}), which we refer to as $\sigma_{\rm full}$,
to a sub-permutation $\sigma_{\rm sub}$, which excludes such exchanges, thus reducing $N_{\rm perm}$ by a factor of 
$2^A$~\cite{Ishii:2006ec, Ishii:2009zr, Aoki:2009ji, Yamazaki:2009ua, Doi:2011gq}.
Second, one finds that since the
color/spinor contractions in the sink are performed prior to evaluating Eq.~(\ref{eq:block}),
$N_{\rm loop}$ is reduced from $6^{2A}\cdot 4^{2A}$ to $6^{A}\cdot 4^{A}$.
Note that the computational cost of evaluating Eqs.~(\ref{eq:fp}) and (\ref{eq:fn}) is negligible,
once up/down quark propagators are determined.
Finally,
this algorithm enables us to reduce 
the computational cost for the momentum projection.
In fact, it is efficient to transform $f^p_\alpha$, $f^n_\beta$ to momentum-space first,
prior to the calculation of Eq.~(\ref{eq:block}).
In this way, (e.g., in the case of the computation of NBS wave functions),
one can perform the zero-momentum projection onto the center of gravity utilizing the convolution technique.
This reduces $N_{\rm vol}$ from $L^{3A}$ down to ${\cal O}(L^{3A-3})$~\cite{Ishii:2009zr}.
Furthermore, if one is interested in only the energy of the system 
using the correlator 
with each sink baryon projected onto zero-momentum (or any fixed momentum), 
$N_{\rm vol} = {\cal O}(1)$, 
insensitive to $A$~\cite{Yamazaki:2009ua}.

We note that an
additional improvement has been carried out in Ref.~\cite{Yamazaki:2009ua},
in the isospin symmetric limit. 
By exploiting the permutation symmetry of protons and neutrons
in the baryon interpolating field as well as other techniques, 
they achieved a significant reduction of $N_{\rm perm}$,
down to $N_{\rm perm} = 93$ for $^3$He ($^3$H) and
$N_{\rm perm} = 1107$ for $^4$He~\cite{Yamazaki:2009ua}.

\section{Unified contraction algorithm}
\label{sec:unified}

We develop a new technique for evaluating contractions in correlation functions
such as those defined in Eq.~(\ref{eq:corr})
by considering the permutation of quarks (Wick contractions) 
and the color/spinor contractions simultaneously.
In doing so, we may
eliminate redundancies as much as possible.
Although the technique is rather general, and may be applied as an extension of either the
straightforward algorithm or block algorithm, we focus our study on the latter case for simplicity.

To demonstrate the idea, we again consider a $2A$--point nucleon correlation function with $A=2$,
represented by block components given in Eq.~(\ref{eq:block}).
In our algorithm, 
we evaluate Eq.~(\ref{eq:block}) under the condition that
quarks of the same flavor have the same space-time source point,
or more generally, have the same space-time smearing function at the source.
%
Under this condition,
a permutation of quark operators in the source is equivalent to 
a permutation of color and spinor indices of the corresponding quark sources,
and thus we can rewrite Eq.~(\ref{eq:block}) as
\begin{eqnarray}
\label{eq:perm-loop1}
\lefteqn{
\Pi_{
\alpha, \beta;\
\alpha',\beta'}
(X_1,X_2;\ X'_1,X'_2)
} \nn
%
%
\label{eq:perm-loop2}
&=&
f_\alpha^p(X_1;\ \xi'_{1}, \xi'_{2}, \xi'_{3}) 
\cdot
f_\beta^n (X_2;\ \xi'_{4}, \xi'_{5}, \xi'_{6}) 
\times C^{pn}_{\alpha' \beta'} (\xi'_1, \cdots, \xi'_6) ,
\end{eqnarray}
where
%
%
%
%
\begin{eqnarray}
\label{eq:list}
\lefteqn{
C^{pn}_{\alpha' \beta'} (\xi'_1, \cdots, \xi'_6)
} \nn
&\equiv&
\sum_{\sigma^{-1}} 
\epsilon_{c_{\sigma(1)}' c_{\sigma(2)}' c_{\sigma(3)}'} (C \Gamma'^{\,p}_1)_{\alpha_{\sigma(1)}',\alpha_{\sigma(2)}'} (\Gamma'^{\,p}_2)_{\alpha',\alpha_{\sigma(3)}'} \nn
&& \quad \times
\epsilon_{c_{\sigma(4)}' c_{\sigma(5)}' c_{\sigma(6)}'} (C \Gamma'^{\,n}_1)_{\alpha_{\sigma(4)}',\alpha_{\sigma(5)}'} (\Gamma'^{\,n}_2)_{\beta',\alpha_{\sigma(6)}'} 
\cdot {\rm sign}(\sigma) ,
\end{eqnarray}
and the sum is carried over the inverse permutations $\sigma^{-1}$.
An essential feature of this result is that
the computation of a permutation is absent in Eq.~(\ref{eq:perm-loop2}).
To compensate for this, one must instead perform a permutation calculation to evaluate Eq.~(\ref{eq:list}).
Since the summand in Eq.~(\ref{eq:list}) is independent of the gauge field, however, this calculation
need only be carried out once, and independently of any lattice simulation.
As was the case for Eq.~(\ref{eq:block}), the permutation sum in Eq.~(\ref{eq:list}) may be taken over either the full permutation ($\sigma_{\rm full}$) or over the sub-permutation ($\sigma_{\rm sub}$).
Generally, there is no difference between summing over $\sigma^{-1}$ or $\sigma$ in Eq.~(\ref{eq:list}) in the former case, but in the latter case there is a difference, depending on the particular sub-permutation chosen.
Note that in Eq.~(\ref{eq:list}), although $\xi_i'$ depends on the quark coordinate index at the source, because of the same-source condition imposed on the quark fields, this index is irrelevant in its evaluation.

We note that the evaluation of Eq.~(\ref{eq:list}) amounts to 
preparing a unified index list for Wick and color/spinor loop contractions
in which only non-zero components of the coefficient matrix
$C^{pn}_{\alpha' \beta'} (\xi'_1, \cdots, \xi'_6)$
are tabulated.
In particular, if there exists any redundancy and/or cancellation among contributions in the original contraction,
they are automatically consolidated when constructing the unified index list.
Considering the sparse nature of $\gamma$-matrices and $\epsilon$-tensors together,
it is expected that 
the number of non-zero elements in
the coefficient matrix
is rather small,
resulting in significant speedups in the computation of correlators.%
\footnote{%
In evaluating multi-hadron correlation functions, correlation functions may suffer from round-off error due to a large number of cancellations among contributing terms \cite{Detmold:2008fn}.
By evaluating Eq.~(\ref{eq:list}), a subset of these cancellations are performed exactly using integer arithmetic, resulting in a reduction in round-off errors.
}
Note also that 
it is unnecessary to assume any symmetry between different flavors (e.g., isospin symmetry)
in this algorithm,
since only permutations among quarks of the same flavor are utilized.
Various techniques to improve the signal in correlation functions have been investigated, including the use of all-to-all propagators \cite{Foley:2005ac} and novel smearing methods such as distillation \cite{Peardon:2009gh}.
Generally, the application of our approach in such cases is straight-forward, although the degree of cancellation achieved in evaluating the analog of Eq.~(\ref{eq:list}) could depend heavily on the details of the operator construction.

\section{Efficiency of the unified contraction algorithm}
\label{sec:results}

In order to examine the efficiency of the unified contraction algorithm,
we explicitly evaluate coefficient matrices for typical examples of interest.
Particularly, we study systems composed of octet baryons.
With regards to the explicit spinor structure of a baryon operator at the source,
we consider two choices,
\begin{eqnarray}
\label{eq:std-op}
(\Gamma'_1, \Gamma'_2) &=&
(\gamma_5, 1) , \\
\label{eq:nr-op}
(\Gamma'_1, \Gamma'_2) &=&
(\gamma_5 P_{nr}, P_{nr}), \qquad P_{nr} \equiv (1+\gamma_4)/2 .
\end{eqnarray}
For each multi-baryon correlator,
we employ either Eq.~(\ref{eq:std-op}) or (\ref{eq:nr-op}) 
for all baryon operators at the source, and do not mix the two choices.
The choice of Eq.~(\ref{eq:nr-op}) has been employed in 
recent multi-baryon studies~\cite{Yamazaki:2009ua, Doi:2011gq},
because 
only the upper half components of Dirac spinors survive (working in the Dirac basis for $\gamma$-matrices), thus reducing the number of spinor loop contractions by a factor of $2^A$.%
\footnote{
Note that although the operator has an apparent non-relativistic form,
the states, which are dynamically generated on the lattice, 
are not subject to any non-relativistic approximation.
}
Hereafter, 
we refer to Eqs.~(\ref{eq:std-op}) and (\ref{eq:nr-op}) 
as ``standard'' and ``non-relativistic''
operators, respectively.
Note that the evaluation of Eq.~(\ref{eq:list}) does not depend on the
spinor structure of a baryon operator at the sink, $(\Gamma_1,\Gamma_2)$.

When evaluating Eq.~(\ref{eq:list}), we may in principle consider two choices
for the permutation, $\sigma_{\rm full}$ and $\sigma_{\rm sub}$,
as discussed in Sec.~\ref{subsec:block} and Sec.~\ref{sec:unified}.
While $\sigma_{\rm sub}$ is trivially a better choice
in the block algorithm, 
$\sigma_{\rm full}$ offers an advantage in the unified contraction scheme.
In particular, note that 
the coefficient matrix obtained from $\sigma_{\rm full}$
solely depends on the structure of the source operators,
while, in the case of $\sigma_{\rm sub}$, 
it also depends on the sink baryons implicitly 
through the definition of $\sigma_{\rm sub}$.
Therefore, when one considers coupled channel correlation functions
where source baryons and sink baryons could be different,
the contraction list obtained from $\sigma_{\rm full}$ has broader utility.

When sink baryons in the correlator are specified,
one may exploit any existing inner-quark exchange symmetries at the sink
to explicitly constrain the sum over indices in Eq.~(\ref{eq:perm-loop2}),
and thus further reduce the computational cost of its evaluation.
For example, if a baryon block is antisymmetric under the
exchange of two indices $\xi'_i$ and $\xi'_j$ for $i\ne j$, 
and the coefficient matrix is also
antisymmetric under exchange of the same two indices 
(as guaranteed under the full permutation),
then one may constrain the sums in Eq.~(\ref{eq:perm-loop2}) such that $\xi'_i < \xi'_j$.
For instance, in the case of multi-nucleon systems, 
such considerations will result in a reduction of $2^A$ in total.%
\footnote{
One can apply a similar procedure for the unified contraction list with $\sigma_{\rm sub}$,
but the computational cost in the evaluation of Eq.~(\ref{eq:perm-loop2}) is
equivalent to the cost of the choice with $\sigma_{\rm full}$, 
as is evident from the definition of Eq.~(\ref{eq:list}).
}
Note that this is the same reduction factor that is achieved by using $\sigma_{\rm sub}$ as opposed to $\sigma_{\rm full}$ in the block algorithm.

We carry out the construction of the unified contraction list
using supercomputers, since in the case of a mass number $ A > 2$,
the computational cost is found to be quite large depending on the operators chosen,
naively growing factorially in each quark number (this exponential growth in computational cost
can be eliminated in some cases by exploiting Pauli exclusion, as will be discussed later on).
It is, however, just a one-time investment, and
we intend to make the lists publicly available for future use.
We investigate the utility of our algorithm by considering two-octet baryon systems in the case of $A=2$,%
and multi-nucleon systems, i.e., $^3$H/$^3$He and $^4$He, for $A=3, 4$.
For simplicity, we consider single channel systems in this study,
while extension to coupled channel systems is straightforward.
The computational cost of correlators using the unified contraction algorithm
can be estimated by counting the number of non-zero elements, $N_{\rm list}$,
in the coefficient matrix under the full permutation.
The total number of terms in the contraction
will then be given by $N_{\rm contr} = N_{\rm list}/2^A$ 
after exploiting the inner-quark exchange to explicitly constrain the sums
in Eq.~(\ref{eq:perm-loop2}).
The values we obtain for $N_{\rm list}$ and $N_{\rm contr}$ are compiled in~\ref{sec:tabs},
together with $N_{\rm perm}$ and  $N_{\rm loop}$ obtained in the block algorithms.

In order to make a comparison of methods easier,
we present an effective number of permutations, defined by
$N^{\rm eff}_{\rm perm} \equiv N_{\rm contr} / N_{\rm loop}$
where $N_{\rm loop}$ is taken from the corresponding block algorithm.
The efficiency of our approach in comparison with the block algorithm
(i.e., the speed-up factor)
is given by 
the ratio $\eta \equiv N_{\rm perm} / N^{\rm eff}_{\rm perm}$,
with a ratio larger than unity indicating an improvement.
From the tables in~\ref{sec:tabs}, one immediately observes that the unified contraction algorithm
yields better efficiency
for all multi-baryon systems under consideration.
What is particularly noteworthy is
the gain for $A=3$ and $4$ with non-relativistic operators.
In the case of $^3$H/$^3$He,
a factor of 192 improvement over the block algorithm is achieved,
and in the case of $^4$He, a factor of 20736 improvement is achieved.
We have checked that these improvement factors are nearly realized
in the actual lattice simulation code, as well.
We also observe significant improvements in the case of $A=2$.
These improvements 
are useful for, e.g., coupled channel calculations 
for YN, YY interactions,
where considerable computational cost would be 
required using the block algorithm~\cite{Sasaki:2011AAA}.

It is in order that we remark on several aspects of 
the unified contraction algorithm results presented in~\ref{sec:tabs}.
%
First, a special property is observed for $^4$He correlators when non-relativistic operators are considered.
Specifically, 
one finds that $N_{\rm list} = 518400$
is exactly the same as $N_{\rm perm}$ obtained in the block algorithm with $\sigma_{\rm full}$.
This can be understood intuitively, by 
noting that color/spinor DoF are completely saturated in $^4$He
when quark sources are taken to be equal.
This is a simple statement that, for every baryon spin component, the $^4$He analog of the coefficient matrix
defined in Eq.~(\ref{eq:list}) is proportional to the product of two epsilon tensors in $\xi'_i$
(one for each flavor).
Similar saturation is realized for $^8$Be, 
when we employ the operator which uses all four spinors of quarks. 
For larger A, such fermion saturation can be exploited to reduce the computational cost of evaluating
the unified contraction list by noting that the coefficient matrix must be proportional to
an epsilon tensor for a subset of indices corresponding to the fermions for which the color/spinor degrees of freedom
are fully saturated.

Second, although we tabulated $N_{\rm contr}$ 
for all possible (upper) baryon spin indices $(\alpha',\beta',\cdots)$ at the source,
it is not always necessary to calculate the correlator for all of them.
For instance, the correlator for $^4$He 
with $\alpha' = \beta'$ or $\gamma' = \delta'$
should be trivially zero because of the anti-commuting property of source baryons.
It is interesting that the unified contraction algorithm 
exposes this feature explicitly, as is evident from the tables in~\ref{sec:tabs}.
In the same way, the $^4$He correlator with, e.g.,
$(\alpha',\beta',\gamma',\delta')$ and $(\beta',\alpha',\delta',\gamma')$ 
should be same, so one can save computational time 
by calculating only one of them.
Third, by imposing additional constraints on baryon sink operators, 
further reduction in computational cost is possible.
For instance, when we consider the correlation function 
with each sink baryon projected onto zero-momentum,
one may exploit any existing exchange symmetry among the same baryons in the sink~\cite{Yamazaki:2009ua}.
This leads to a further reduction of the computational cost,
e.g., by a factor of two and four for $^3$H/$^3$He and $^4$He, respectively.
Depending of the system of concern,
one may also exploit additional symmetries, if any, 
to skip the computation of redundant correlators.
For example, if one assumes isospin symmetry, the computational cost
of the above mentioned $^4$He correlator can be reduced further by about a factor of two.

The proposed algorithm is rather general,
and so it is possible to extend the technique to other systems of interest.
An immediate application is to multi-hadron correlators
including mesons, although for purely mesonic correlators,
it remains to be determined whether the technique offers any advantage over the recursive approaches of Refs.~\cite{Detmold:2010au,Detmold:2012wc}.
Furthermore, the method may be extended to various problems
in quantum mechanics, where Slater determinants often appear as the subject of interest.
If the coefficients of the Slater determinants
have a sparse nature, a similar prescription
may provide an efficient alternative.

Finally, let us discuss the limitations of the unified contraction algorithm.
In order to satisfy the same-source condition
described in Sec.~\ref{sec:unified},
the number of quarks associated with each flavor must be
less than or equal to 12 due to Pauli exclusion.
Therefore, the maximum mass number allowed 
is $A_{\rm max}=8$ in 2-flavor space and $A_{\rm max}=12$ in 3-flavor space, respectively.
We note, however, that even for a system with $A > A_{\rm max}$,
the presented algorithm is expected to be efficient 
since it can reduce the computational cost
corresponding to the subspace of permutations,
which is spanned by imposing the same-source condition on as many quarks as possible.
Further studies 
are currently underway.


In the limit of large $A$,
suppressing the computation of Wick contractions in Eq.~(\ref{eq:corr}) becomes most important,
since $N_{\rm perm}$ grows factorially in quark number,
whereas the others ($N_{\rm loop}$, $N_{\rm vol}$) grow exponentially with quark number.
In fact, such an algorithm was proposed in~\cite{Kaplan:DWF10yrsTalk}, where 
the Wick contractions in Eq.~(\ref{eq:corr}) 
are expressed in terms of a determinant of quark propagators.
It is then essential to realize that the computational cost of the determinant of an $n\times n$ matrix
can be reduced from ${\cal O}(n!)$ to ${\cal O}(n^3)$ by employing LU-decomposition.
Unfortunately, this is not an efficient algorithm for light nuclei such as $^4$He,
since $N_{\rm loop}$ and/or $N_{\rm vol}$ remain overwhelmingly large as discussed in Sec.~\ref{subsec:naive}.
However, as the mass number $A$ grows, the determinant algorithm would presumably become a useful approach.

\section{Summary}
\label{sec:summary}

We have proposed an efficient algorithm for calculating multi-baryon correlation functions on the lattice.
By considering the permutation of quarks (Wick contractions) 
and the color/spinor contractions simultaneously, 
we have shown that there exist 
large redundancies in the original contraction.
We have developed a method to construct a unified index list for the contraction 
in which the redundancies are eliminated.
It is noted that an assumption on the symmetry between different flavors (e.g., isospin symmetry)
is not required in this algorithm, although imposing such symmetries leads to further computational savings.
Possible extensions of this algorithm have also been discussed.

In order to determine how efficient the algorithm is,
we have investigated several typical examples of interest,
namely, two-octet baryon systems, $^3$H, $^3$He and $^4$He.
We have found that a significant speedup is achieved
in all cases, in particular,
by a factor of 192 for $^3$H and $^3$He nuclei and 
a factor of 20736 for the $^4$He nucleus.
For typical correlators of concern, where each nucleon is projected onto zero-momentum,
further speedup can be achieved, e.g., by a factor of 2 and 4 for $^3$H/$^3$He and $^4$He, respectively.
This achievement takes a significant step towards the ultimate objective of 
studying nuclear physics from first principles lattice simulations of QCD (+ QED).


\section*{Acknowledgments}

We thank Dr. H.~Suzuki for fruitful discussions; T.D. also thanks colleagues in the HAL QCD Collaboration
for helpful discussions.
The numerical simulations have been performed on 
SR16000 at YITP in Kyoto University, 
Blue Gene/Q at KEK
and
FX10 at Tokyo University.
%
M.G.E. is supported by the Foreign Postdoctoral Researcher Program at RIKEN.
This research is supported in part by 
MEXT Grant-in-Aid for Young Scientists (B) (24740146 and 23740227),
the Large Scale Simulation Program of KEK No.12-11 (FY2011-2012)
and SPIRE (Strategic Program for Innovative REsearch).


\appendix

\section{Computational cost of the unified contraction algorithm}
\label{sec:tabs}

Here, we show the computational cost of the unified contraction algorithm for various multi-baryon correlators.
In particular, we tabulate the number of non-zero entries ($N_{\rm list}$) appearing in the coefficient matrix,
the total number of contractions ($N_{\rm contr}$),
and  the effective permutation number
($N_{\rm perm}^{\rm eff} \equiv N_{\rm contr}/N_{\rm loop}$).
As discussed in Sec.~\ref{sec:unified}, the number of contractions required to compute the correlation function is given by
$N_{\rm contr} = N_{\rm list}/2^A$ in the case of nucleons.
For comparison, we also provide
$N_{\rm perm}$ and $N_{\rm loop}$ for the block algorithm.
The increase in efficiency achieved by the unified contraction algorithm
over the block algorithm is given by the ratio 
$\eta \equiv N_{\rm perm} / N_{\rm perm}^{\rm eff}$,
where $N_{\rm perm}$ is the number of permutations required
in the block algorithm using the optimal choice, $\sigma_{\rm sub}$.

Tables are provided for single channel two-octet baryon systems, $^3$H/$^3$He and $^4$He.
For each baryon in the system, only the upper spinor components are considered,
i.e., $\alpha_i' = (0,1)$ for $i=1,\ldots,A$, since
each baryon field is expected to couple strongly to a corresponding positive parity baryon.
Because the system is symmetric under the flip of all spin indices,
we only give results for 
$\alpha_1' = 0$. 
For baryon interpolating fields in the source,
we consider both standard and non-relativistic operators, 
as described in Sec.~\ref{sec:results}.

\subsection{Two-octet baryon systems}
\label{subsec:tabs:2b}

We consider single channel systems of two-octet baryons without assuming flavor symmetry.
The contraction list falls into three classes based on the flavor content of baryons:
(i) $pp$, $nn$, $\Sigma^+\Sigma^+$, $\Sigma^-\Sigma^-$, $\Xi^0\Xi^0$, $\Xi^-\Xi^-$ systems,
(ii) $pn$, $\Sigma^+\Xi^0$, $\Sigma^-\Xi^-$ systems and
(iii) $p\Sigma^+$, $n\Sigma^-$, $\Xi^0\Xi^-$ systems.
The correlation function under consideration is given by Eq.~(\ref{eq:pn}), or an analog of it.


\begin{table}[H] 
\label{tab:pp}
\caption{$pp$, $nn$, $\Sigma^+\Sigma^+$, $\Sigma^-\Sigma^-$, $\Xi^0\Xi^0$, $\Xi^-\Xi^-$ systems with the standard operators.}
\begin{center}
\begin{tabular}{c|ccc|ccc|c}
\hline\hline
source spin &  \multicolumn{3}{|c|}{block algorithm}   & \multicolumn{3}{|c|}{unified contraction algorithm} & efficiency \\ \hline
$(\alpha',\beta')$
             & $N_{\rm loop}$  & $N_{\rm perm}$\,\scsize{($\sigma_{\rm full}$)} 
                                    & $N_{\rm perm}$\,\scsize{($\sigma_{\rm sub}$)} & $N_{\rm list}$ & $N_{\rm contr}$ & $N_{\rm perm}^{\rm eff}$ & $\eta$ \\ \cline{1-8}
(0, 0) &  576       &  48                & 12                &  0           &  0       & 0      & -   \\
(0, 1) &  576       &  48                & 12                &  11088       &  2772    & 4.8    & 2.5 \\ \hline\hline
\end{tabular}
\end{center}
\caption{Same as above, but with the non-relativistic operators.}
\begin{center}
\begin{tabular}{c|ccc|ccc|c}
\hline\hline
source spin &  \multicolumn{3}{|c|}{block algorithm}   & \multicolumn{3}{|c|}{unified contraction algorithm} & efficiency \\ \hline
$(\alpha',\beta')$
             & $N_{\rm loop}$  & $N_{\rm perm}$\,\scsize{($\sigma_{\rm full}$)} 
                                    & $N_{\rm perm}$\,\scsize{($\sigma_{\rm sub}$)} & $N_{\rm list}$ & $N_{\rm contr}$ & $N_{\rm perm}^{\rm eff}$ & $\eta$ \\ \cline{1-8}
(0, 0) &  144       &  48                & 12                &  0       &  0         & 0      & -          \\
(0, 1) &  144       &  48                & 12                &  1008    &  252       & 1.8    & 6.9        \\ \hline\hline
\end{tabular}
\end{center}
\end{table}


\begin{table}[H] 
\label{tab:pn}
\caption{$pn$, $\Sigma^+\Xi^0$, $\Sigma^-\Xi^-$ systems with the standard operators.}
\begin{center}
\begin{tabular}{c|ccc|ccc|c}
\hline\hline
source spin &  \multicolumn{3}{|c|}{block algorithm}   & \multicolumn{3}{|c|}{unified contraction algorithm} & efficiency \\ \hline
$(\alpha',\beta')$
             & $N_{\rm loop}$  & $N_{\rm perm}$\,\scsize{($\sigma_{\rm full}$)} 
                                    & $N_{\rm perm}$\,\scsize{($\sigma_{\rm sub}$)} & $N_{\rm list}$ & $N_{\rm contr}$ & $N_{\rm perm}^{\rm eff}$ & $\eta$ \\ \cline{1-8}
(0, 0) &  576       &  36                &  9                &  8316    &  2079       & 3.6    & 2.5              \\
(0, 1) &  576       &  36                &  9                &  9432    &  2358       & 4.1    & 2.2              \\ \hline\hline
\end{tabular}
\end{center}
\caption{Same as above, but with the non-relativistic operators.}
\begin{center}
\begin{tabular}{c|ccc|ccc|c}
\hline\hline
source spin &  \multicolumn{3}{|c|}{block algorithm}   & \multicolumn{3}{|c|}{unified contraction algorithm} & efficiency \\ \hline
$(\alpha',\beta')$
             & $N_{\rm loop}$  & $N_{\rm perm}$\,\scsize{($\sigma_{\rm full}$)} 
                                    & $N_{\rm perm}$\,\scsize{($\sigma_{\rm sub}$)} & $N_{\rm list}$ & $N_{\rm contr}$ & $N_{\rm perm}^{\rm eff}$ & $\eta$ \\ \cline{1-8}
(0, 0) &  144       &  36                &  9                &   756   &   189         & 1.3 & 6.9  \\
(0, 1) &  144       &  36                &  9                &  1008   &   252         & 1.8 & 5.1  \\ \hline\hline
\end{tabular}
\end{center}
\end{table}


\begin{table}[H] 
\label{tab:psig}
\caption{$p\Sigma^+$, $n\Sigma^-$, $\Xi^0\Xi^-$ systems with the standard operators.}
\begin{center}
\begin{tabular}{c|ccc|ccc|c}
\hline\hline
source spin &  \multicolumn{3}{|c|}{block algorithm}   & \multicolumn{3}{|c|}{unified contraction algorithm} & efficiency \\ \hline
$(\alpha',\beta')$
             & $N_{\rm loop}$  & $N_{\rm perm}$\,\scsize{($\sigma_{\rm full}$)} 
                                    & $N_{\rm perm}$\,\scsize{($\sigma_{\rm sub}$)} & $N_{\rm list}$ & $N_{\rm contr}$ & $N_{\rm perm}^{\rm eff}$ & $\eta$ \\ \cline{1-8}
(0, 0) &  576       &  24                & 6                 &  5400       &  1350           & 2.3    & 2.6 \\
(0, 1) &  576       &  24                & 6                 &  7776       &  1944           & 3.4    & 1.8 \\ \hline\hline
\end{tabular}
\end{center}
\caption{Same as above, but with the non-relativistic operators.}
\begin{center}
\begin{tabular}{c|ccc|ccc|c}
\hline\hline
source spin &  \multicolumn{3}{|c|}{block algorithm}   & \multicolumn{3}{|c|}{unified contraction algorithm} & efficiency \\ \hline
$(\alpha',\beta')$
             & $N_{\rm loop}$  & $N_{\rm perm}$\,\scsize{($\sigma_{\rm full}$)} 
                                    & $N_{\rm perm}$\,\scsize{($\sigma_{\rm sub}$)} & $N_{\rm list}$ & $N_{\rm contr}$ & $N_{\rm perm}^{\rm eff}$ & $\eta$ \\ \cline{1-8}
(0, 0) &  144       &  24                & 6                 &  648   &  162         & 1.1  & 5.3 \\
(0, 1) &  144       &  24                & 6                 &  864   &  216         & 1.5  & 4.0 \\ \hline\hline
\end{tabular}
\end{center}
\end{table}

\subsection{Three-octet baryon systems}
\label{subsec:tabs:3N}

We consider the correlation function for $^3$H given by
\begin{eqnarray}
\label{eq:pnn}
\Pi_{\alpha\beta\gamma;\, \alpha'\beta'\gamma'} 
&\equiv& 
\langle
p_\alpha n_\beta n_\gamma \
\bar n'_{\gamma'} \bar n'_{\beta'} \bar p'_{\alpha'}
\rangle . 
\end{eqnarray}
$^3$He, and the hyperon analogs
$\Sigma^+\Sigma^+\Xi^0$,
$\Sigma^-\Sigma^-\Xi^-$,
$\Sigma^+\Xi^0\Xi^0$, and
$\Sigma^-\Xi^-\Xi^-$
share the same contraction list.

\begin{table}[H] 
\label{tab:3N}
\caption{$^3$H, $^3$He systems with the standard operators.}
\begin{center}
\begin{tabular}{c|ccc|ccc|c}
\hline\hline
source spin &  \multicolumn{3}{|c|}{block algorithm}   & \multicolumn{3}{|c|}{unified contraction algorithm} & efficiency \\ \hline
$(\alpha',
  \beta',
  \gamma')$
             & $N_{\rm loop}$  & $N_{\rm perm}$\,\scsize{($\sigma_{\rm full}$)} & $N_{\rm perm}$\,\scsize{($\sigma_{\rm sub}$)}
                                                               & $N_{\rm list}$ & $N_{\rm contr}$ & $N_{\rm perm}^{\rm eff}$ & $\eta$ \\ \cline{1-8}
(0, 0, 0)    &  13824     & 2880                & 360                &  0          &  0               & 0     & -  \\
(0, 0, 1)    &  13824     & 2880                & 360                &  3775680    &  471960          & 34.1  & 10.5 \\
(0, 1, 0)    &  13824     & 2880                & 360                &  3775680    &  471960          & 34.1  & 10.5  \\
(0, 1, 1)    &  13824     & 2880                & 360                &  0          &  0               & 0     & -  \\ \hline\hline
\end{tabular}
\end{center}
\caption{Same as above, but with the non-relativistic operators.}
\begin{center}
\begin{tabular}{c|ccc|ccc|c}
\hline\hline
source spin &  \multicolumn{3}{|c|}{block algorithm}   & \multicolumn{3}{|c|}{unified contraction algorithm} & efficiency \\ \hline
$(\alpha',
  \beta',
  \gamma')$
             & $N_{\rm loop}$  & $N_{\rm perm}$\,\scsize{($\sigma_{\rm full}$)} & $N_{\rm perm}$\,\scsize{($\sigma_{\rm sub}$)}
                                                               & $N_{\rm list}$ & $N_{\rm contr}$ & $N_{\rm perm}^{\rm eff}$ & $\eta$ \\ \cline{1-8}
(0, 0, 0)    &  1728      & 2880                & 360                &  0          &  0               & 0      & -   \\
(0, 0, 1)    &  1728      & 2880                & 360                &  25920      &  3240            & 1.9    & 192 \\
(0, 1, 0)    &  1728      & 2880                & 360                &  25920      &  3240            & 1.9    & 192 \\
(0, 1, 1)    &  1728      & 2880                & 360                &  0          &  0               & 0      & -   \\ \hline\hline
\end{tabular}
\end{center}
\end{table}

\subsection{Four-octet baryon systems}
\label{subsec:tabs:4N}

We consider the correlation function of $^4$He given by
\begin{eqnarray}
\label{eq:ppnn}
%
%
\Pi_{\alpha\beta\gamma\delta;\, \alpha'\beta'\gamma'\delta'} 
%
&\equiv& 
\langle
p_\alpha p_\beta n_\gamma n_\delta \
\bar n'_{\delta'} \bar n'_{\gamma'} \bar p'_{\beta'} \bar p'_{\alpha'}
\rangle .
\end{eqnarray}
The hyperon analogs $\Sigma^+\Sigma^+\Xi^0\Xi^0$ and $\Sigma^-\Sigma^-\Xi^-\Xi^-$
share the same contraction list.

\begin{table}[H] 
\label{tab:4N}
\caption{$^4$He system with the standard operators.}
\begin{center}
\begin{tabular}{c|ccc|ccc|c}
\hline\hline
source spin &  \multicolumn{3}{|c|}{block algorithm}   & \multicolumn{3}{|c|}{unified contraction algorithm} & efficiency \\ \hline
$(\alpha',
  \beta',
  \gamma',
  \delta')$ 
             & $N_{\rm loop}$  & $N_{\rm perm}$\,\scsize{($\sigma_{\rm full}$)} & $N_{\rm perm}$\,\scsize{($\sigma_{\rm sub}$)}
                                                               & $N_{\rm list}$ & $N_{\rm contr}$ & $N_{\rm perm}^{\rm eff}$ & $\eta$ \\ \cline{1-8}
(0, 0, 0, 0) &  331776    & 518400              & 32400              &  0             &  0             & 0       & - \\
(0, 0, 0, 1) &  331776    & 518400              & 32400              &  0             &  0             & 0       & - \\
(0, 0, 1, 0) &  331776    & 518400              & 32400              &  0             &  0             & 0       & - \\
(0, 0, 1, 1) &  331776    & 518400              & 32400              &  0             &  0             & 0       & - \\
(0, 1, 0, 0) &  331776    & 518400              & 32400              &  0             &  0             & 0       & - \\
(0, 1, 0, 1) &  331776    & 518400              & 32400              &  1407974400    &  87998400      & 265.2   & 122.2 \\
(0, 1, 1, 0) &  331776    & 518400              & 32400              &  1407974400    &  87998400      & 265.2   & 122.2 \\
(0, 1, 1, 1) &  331776    & 518400              & 32400              &  0             &  0             & 0       & - \\ \hline\hline
\end{tabular}
\end{center}
\caption{Same as above, but with the non-relativistic operators.}
\begin{center}
\begin{tabular}{c|ccc|ccc|c}
\hline\hline
source spin &  \multicolumn{3}{|c|}{block algorithm}   & \multicolumn{3}{|c|}{unified contraction algorithm} & efficiency \\ \hline
$(\alpha',
  \beta',
  \gamma',
  \delta')$ 
             & $N_{\rm loop}$  & $N_{\rm perm}$\,\scsize{($\sigma_{\rm full}$)} & $N_{\rm perm}$\,\scsize{($\sigma_{\rm sub}$)}
                                                               & $N_{\rm list}$ & $N_{\rm contr}$ & $N_{\rm perm}^{\rm eff}$ & $\eta$ \\ \cline{1-8}
(0, 0, 0, 0) &  20736     & 518400      & 32400              &  0           & 0      & 0     & -  \\
(0, 0, 0, 1) &  20736     & 518400      & 32400              &  0           & 0      & 0     & -  \\
(0, 0, 1, 0) &  20736     & 518400      & 32400              &  0           & 0      & 0     & -  \\
(0, 0, 1, 1) &  20736     & 518400      & 32400              &  0           & 0      & 0     & -  \\
(0, 1, 0, 0) &  20736     & 518400      & 32400              &  0           & 0      & 0     & -  \\
(0, 1, 0, 1) &  20736     & 518400      & 32400              &  518400      & 32400  & 1.6   & 20736  \\
(0, 1, 1, 0) &  20736     & 518400      & 32400              &  518400      & 32400  & 1.6   & 20736  \\
(0, 1, 1, 1) &  20736     & 518400      & 32400              &  0           & 0      & 0     & -  \\ \hline \hline
\end{tabular}
\end{center}
\end{table}




\begin{thebibliography}{00}


\bibitem{Ishii:2006ec}
  N.~Ishii, S.~Aoki and T.~Hatsuda,
  Phys.\ Rev.\ Lett.\  {\bf 99}, 022001 (2007)
  [nucl-th/0611096].

\bibitem{Ishii:2009zr} 
  N.~Ishii, S.~Aoki and T.~Hatsuda,
  PoS LATTICE {\bf 2008}, 155 (2008)
  [arXiv:0903.5497 [hep-lat]].

\bibitem{Aoki:2009ji}
  S.~Aoki, T.~Hatsuda and N.~Ishii,
  Prog.\ Theor.\ Phys.\  {\bf 123}, 89 (2010)
  [arXiv:0909.5585 [hep-lat]].


\bibitem{Yamazaki:2009ua}
  T.~Yamazaki, Y.~Kuramashi and A.~Ukawa, [PACS-CS Collaboration],
  Phys.\ Rev.\  {\bf D81}, 111504 (2010).
  [arXiv:0912.1383 [hep-lat]].


\bibitem{Doi:2011gq} 
  T.~Doi, {\it et al.}  [HAL QCD Collaboration],
  Prog.\ Theor.\ Phys.\  {\bf 127}, 723 (2012)
  [arXiv:1106.2276 [hep-lat]].


\bibitem{Lepage:1989hd}
  G.~P.~Lepage,
  in {\it From Actions to Answers: Proceedings of the TASI 1989},
  edited by T.~Degrand and D.~Toussaint
  (World Scientific, Singapore, 1990).

\bibitem{Luscher:1990ck}
  M.~Luscher and U.~Wolff,
  Nucl.\ Phys.\  B {\bf 339} (1990) 222.

\bibitem{Fleming:2004hs} 
  G.~T.~Fleming,
  hep-lat/0403023.

\bibitem{Beane:2009kya}
  S.~R.~Beane {\it et al.},
  Phys.\ Rev.\  D {\bf 79}, 114502 (2009)
  [arXiv:0903.2990 [hep-lat]].

\bibitem{Beane:2009gs} 
  S.~R.~Beane, {\it et al.},
  Phys.\ Rev.\ D {\bf 80}, 074501 (2009)
  [arXiv:0905.0466 [hep-lat]].

\bibitem{Yamazaki:2011nd} 
  T.~Yamazaki, Y.~Kuramashi and A.~Ukawa,
  Phys.\ Rev.\ D {\bf 84}, 054506 (2011)
  [arXiv:1105.1418 [hep-lat]].

\bibitem{Ishii:2012aa} 
  N.~Ishii, {\it et al.}  [HAL QCD Collaboration],
  Phys.\ Lett.\  {\bf B}, in press,
  arXiv:1203.3642 [hep-lat].



\bibitem{Fukugita:1994na} 
  M.~Fukugita, Y.~Kuramashi, H.~Mino, M.~Okawa and A.~Ukawa,
  Phys.\ Rev.\ Lett.\  {\bf 73}, 2176 (1994)
  [hep-lat/9407012].

\bibitem{Fukugita:1994ve} 
  M.~Fukugita, Y.~Kuramashi, M.~Okawa, H.~Mino and A.~Ukawa,
  Phys.\ Rev.\ D {\bf 52}, 3003 (1995)
  [hep-lat/9501024].

\bibitem{Luscher:1985dn} 
  M.~Luscher,
  Commun.\ Math.\ Phys.\  {\bf 104}, 177 (1986).

\bibitem{Luscher:1986pf} 
  M.~Luscher,
  Commun.\ Math.\ Phys.\  {\bf 105}, 153 (1986).

\bibitem{Luscher:1990ux}
  M.~Luscher,
  Nucl.\ Phys.\  B {\bf 354}, 531 (1991).

\bibitem{Beane:2011iw} 
  S.~R.~Beane {\it et al.}  [NPLQCD Collaboration],
  Phys.\ Rev.\ D {\bf 85}, 054511 (2012)
  [arXiv:1109.2889 [hep-lat]].


\bibitem{Nemura:2008sp}
  H.~Nemura, N.~Ishii, S.~Aoki and T.~Hatsuda,
  Phys.\ Lett.\  {\bf B673}, 136 (2009)
  [arXiv:0806.1094 [nucl-th]].

\bibitem{Inoue:2010hs}
  T.~Inoue {\it et al.} [HAL QCD Collaboration],
  Prog.\ Theor.\ Phys.\  {\bf 124}, 591 (2010)
  [arXiv:1007.3559 [hep-lat]].

\bibitem{Inoue:2010es}
  T.~Inoue {\it et al.} [HAL QCD Collaboration],
  Phys.\ Rev.\ Lett.\  {\bf 106}, 162002 (2011)
  [arXiv:1012.5928 [hep-lat]].

\bibitem{Murano:2011nz}
  K.~Murano, N.~Ishii, S.~Aoki and T.~Hatsuda,
  Prog.\ Theor.\ Phys.\  {\bf 125}, 1225 (2011)
  [arXiv:1103.0619 [hep-lat]].

\bibitem{Aoki:2011gt} 
  S.~Aoki {\it et al.}  [HAL QCD Collaboration],
  Proc.\ Japan Acad.\ B {\bf 87}, 509 (2011)
  [arXiv:1106.2281 [hep-lat]].

\bibitem{Inoue:2011ai} 
  T.~Inoue {\it et al.}  [HAL QCD Collaboration],
  Nucl.\ Phys.\  A {\bf 881}, 28 (2012)
  arXiv:1112.5926 [hep-lat].

\bibitem{Sasaki:2011AAA}
  K.~Sasaki [HAL QCD Collaboration],
  PoS LATTICE {\bf 2011}, 173 (2011).

\bibitem{Ikeda:2011qm}
  Y.~Ikeda  [HAL QCD Collaboration],
  PoS LATTICE {\bf 2011}, 159 (2011)
  arXiv:1111.2663 [hep-lat].


\bibitem{Detmold:2008fn} 
  W.~Detmold, M.~J.~Savage, A.~Torok, S.~R.~Beane, T.~C.~Luu, K.~Orginos and A.~Parreno,
  Phys.\ Rev.\ D {\bf 78}, 014507 (2008)
  [arXiv:0803.2728 [hep-lat]].

\bibitem{Foley:2005ac}
 J.~Foley, K.~Jimmy Juge, A.~O'Cais, M.~Peardon, S.~M.~Ryan and J.~-I.~Skullerud,
 Comput.\ Phys.\ Commun.\  {\bf 172}, 145 (2005)
 [hep-lat/0505023].

\bibitem{Peardon:2009gh}
  M.~Peardon {\it et al.}  [Hadron Spectrum Collaboration],
  Phys.\ Rev.\ D {\bf 80}, 054506 (2009)
  [arXiv:0905.2160 [hep-lat]].

\bibitem{Detmold:2010au}
  W.~Detmold and M.~J.~Savage,
  Phys.\ Rev.\  D {\bf 82}, 014511 (2010)
  [arXiv:1001.2768 [hep-lat]].

\bibitem{Detmold:2012wc} 
  W.~Detmold, K.~Orginos and Z.~Shi,
  arXiv:1205.4224 [hep-lat].

\bibitem{Kaplan:DWF10yrsTalk}
  D.~Kaplan, Talk given at Workshop on ``Domain Wall Fermions at Ten Years'',
  Mar. 2007, RIKEN BNL Research Center, NY, USA,
  \verb|https://www.bnl.gov/riken/dwf/talks/files/Kaplan_BNL_talk.pdf|




\end{thebibliography}



\end{document}